\documentstyle[pra,aps,twocolumn]{revtex}

\begin{document}

\title
{
Microscopic theory of the interaction of ultracold dense Bose and Fermi
gases with electromagnetic field
}

\author
{K.V.Krutitsky
\thanks{
		 Permanent address:
		 Ulyanovsk Branch of Moscow Institute of Radio
		 Engineering
		 and Electronics of Russian Academy of Sciences,
		 P.B.9868,
		 48, Goncharov Str., Ulyanovsk 432011, Russia;
		 e-mail: ufire@mv.ru
},
F.Burgbacher
and J.Audretsch
\thanks{e-mail: Juergen.Audretsch@uni-konstanz.de}
\\
Fakult\"at f\"ur Physik,
Universit\"at Konstanz, Fach M 674,
D-78457 Konstanz, Germany
}

\date{2d July, 1999}

\maketitle

\begin{abstract}
We present the rigorous microscopic
quantum theory of the interaction of ultracold Bose and Fermi gases with the
electromagnetic field of vacuum and laser photons. The main attention
has been paid to the consistent consideration of dynamical dipole-dipole
interactions.
The theory developed is shown to be consistent with the general principles
of the canonical quantization of electromagnetic field in a medium.
Starting from the first principles of QED we have derived the general
system of Maxwell-Bloch
equations for atomic creation and annihilation operators and the propagation
equation for the laser field which can be used
for the self-consistent
analysis of various linear and nonlinear phenomena in atom optics at high
densities of the atomic system.
All known equations which are used for the description
of the behaviour of an ultracold atomic ensemble in a radiation field
can be obtained from our general system of
equations in a low-density limit.
\end{abstract}


\section{Introduction}
~~~

   In the last decade  rapid progress is achieved in laser cooling
of neutral atoms. This leads to the vigorous development of the new
field of atom optics. One can distinguish  linear
and nonlinear effects in atom optics. Theoretical analysis of the linear
effects in atom optics is carried out without consideration of
atomic interactions. On the other hand, it
was shown that the consideration of dynamical dipole-dipole
interactions in the radiation field leads to the nonlinear effects
such as a nonlinear Bragg diffraction and the
propagation of atomic solitons. That is why in recent years the problem
of the interaction of ultracold atoms in the field of external laser
radiation has attracted a lot of attention.

   The analysis carried out by O.Morice et al.\cite{MOR95}
and J.Ruostekoski and J.Javanainen \cite{RJ,RJcm} was mainly concentrated
on the
investigation of the properties of the laser radiation modified by atomic
dipole-dipole interactions. It was shown that this can be described
by the refractive index which is governed by the Clausuius-Mossotti
relation known from classical optics if we neglect quantum correlations.
If the quantum correlations caused by Bose or Fermi statistics are taken
into account, the formula for the refractive index contains additional
terms defined by a position dependent correlation
function\cite{MOR95,RJ,RJcm}.

The modification of the properties of the laser radiation should have a
back influence on the behavior of ultracold atomic ensemble (for example,
the motion of atomic beam). An attempt to consider this back
influence was undertaken by several authors
\cite{RJ,LEN93,ZHA94a,LEN94,CAS95,WAL97}.
In order to incorporate two-body interactions G.Lenz, P.Meystre, and
E.M.Wright proceeded by eliminating the transverse vacuum field and replacing
it by the phenomenological contact potential\cite{LEN93}.
Dynamical
dipole-dipole interactions which play more important role than contact
interactions were taken into account in another
papers\cite{RJ,ZHA94a,LEN94,CAS95,WAL97}.
However, in papers by W.Zhang and D.Walls \cite{ZHA94a}
and G.Lenz et al.\cite{LEN94}
the averaged polarization of
ultracold atomic ensemble was computed as a function of the incident
laser field, whereas it should be a function of the macroscopic or the
local
field which are different from the external laser field due to dynamical
dipole-dipole interactions. Wallis \cite{WAL97} paid attention
to this fact and  used the correct form of the equations for
the electromagnetic field. However, his equation for the matter field is
written in the lower order of the density in comparison with the equation
for the electromagnetic field \cite{KBA99}. This problem was considered also
by Y.Castin and K.M{\o}lmer \cite{CAS95} and
J.Ruostekoski and J.Javanainen \cite{RJ}.
But the equations which are used in these
papers are very complicated because they are written down in terms
of the local field and dipole-dipole interactions are presented explicitly
in the form of the sum over dipole fields. This makes the analysis
very difficult and the equations, which govern the time evolution
of the matter fields, remained unsolved.

In our recent paper \cite{KBA99} we have developed the rigorous quantum
theory of the
interaction of an ultracold atomic ensemble with the electromagnetic
field of vacuum and laser photons.
The main attention has thereby  been paid to the consistent consideration of dynamical
dipole-dipole interactions. We have shown that the retardation effects
influence significantly the behaviour of atomic ensemble in the
radiation field. Making use of the Lorentz-Lorenz relation, we have
obtained the general system of equations for
atomic creation and annihilation operators and the propagation equation
for the laser field which can be used for the self-consistent analysis
of various linear and nonlinear phenomena in atom optics at high
densities of the atomic system.
We have shown that all equations which are used
up to now for the description of an ultracold atomic ensemble
subject to the influence of the laser radiation can be obtained
from our general system of equations by series expansion with
respect to the density.

In spite of the fact that most of the results obtained in \cite{KBA99}
remain valid in the case of Fermi-Dirac gases, we considered explicitly
only the case of Bose-Einstein statistics. In addition, our treatment
of the electromagnetic field was not consistent with the QED canonical
quantization procedure. The vacuum fluctuations of the photon field
were separated from the electromagnetic field operator. Such an operation
leads to the violation of the QED canonical commutation relations.
The aim of the present paper is to fill these gaps.

\section{The model and Hamiltonian}
~~~

We discuss the combined system of a gas of identical neutral
polarizable atoms interacting with an incident monochromatic field
${\bf E}_{in}({\bf r},t)$ with frequency $\omega_L$. The probe field
${\bf E}_{in}$ is a classical external field. No operator is related with it.
It is fixed and therefore not subject to a dynamics. For example, it may be
a standing or running laser field. The quantized center-of-mass motion of
the atoms is taken into account. The atoms are represented as point-like
neutral two-level atoms with states $|1\rangle$ and $|2\rangle$ with ``bare"
transition frequency $\omega_a$. The atoms interact via the resonant
dipole-dipole interaction of the induced point dipoles. This interaction is
mediated by the exchange of {\em photons} with annihilation and creation
operators $\hat{c}_{{\bf k}\lambda}$ and $\hat{c}^\dagger_{{\bf k}\lambda}$
referring in lowest order to plane waves in vacuum with wave vector ${\bf k}$
and polarization $\lambda$.

To take into account quantum statistical effects for the two-level atoms, we
turn to quantized matter fields represented by two-component vectors
\begin{equation}
\psi({\bf r},t) =
\hat\psi_1({\bf r},t)|1\rangle+
\hat\psi_2({\bf r},t)|2\rangle
\;.
\nonumber
\end{equation}
The equal time commutation relations are
\begin{eqnarray}
\label{komm-equal}
\left[
     \hat\psi_i({\bf r}) ,
     \hat\psi_j({\bf r}')
\right]_q
&=&
\left[
    \hat{\psi}_i^\dagger({\bf r}) ,
    \hat{\psi}_j^\dagger({\bf r}')
\right]_q
=0
\;,
\nonumber\\
\left[
     \hat\psi_i({\bf r}) ,
     \hat\psi_j^\dagger({\bf r}')
\right]_q
&=&
\delta_{ij} \delta\left( {\bf r}-{\bf r}' \right)
\;,
\nonumber\\
&&i,j=1,2
\;,\quad
q=c,a,
\end{eqnarray}
where $[\hat A,\hat B]_c$ is the commutator of the operators
$\hat A$ and $\hat B$,
and $[\hat A,\hat B]_a$ is the anticommutator, which corresponds
to Bose-Einstein or Fermi-Dirac statistics, respectively.
In terms of the operators $\hat \psi_1$ and $\hat \psi_2$
and in the multipolar formulation of QED~\cite{COH89,KBA99}
the Hamiltonian operator takes then the form:
\begin{eqnarray}
\label{ham-second-quant}
\hat {\cal H}
&=&
\hat {\cal H}_A + \hat H_F + \hat {\cal H}_{AI} + \hat {\cal H}_{AF}
\;,
\\
\hat {\cal H}_A
&=&
\sum_{j=1}^2
\int d{\bf r}
\hat{\psi}_j^\dagger({\bf r},t)
\left(
     - \frac{\hbar^2 \nabla^2}{2m}
\right)
\hat{\psi}_j({\bf r},t)
\nonumber\\
&&+
\int d{\bf r} \hat{\psi}_2^\dagger({\bf r},t)
\hbar \omega_a \hat{\psi}_2({\bf r},t)
\;,
\nonumber\\
\hat H_F
&=&
\sum_{{\bf k}\lambda}
\hbar \omega_k
\hat c^{\dagger}_{{\bf k}\lambda}(t)
\hat{c}_{{\bf k}\lambda}(t)
\;,
\nonumber\\
\hat {\cal H}_{AI}
&=&
-
\int d{\bf r}
\hat {\bf P}({\bf r},t) {\bf E}_{in}({\bf r},t)
\;,
\nonumber\\
\hat {\cal H}_{AF}
&=&
-
\int d{\bf r}
\hat {\bf P}({\bf r},t)
\hat {\bf D}_{mic}({\bf r},t)
\;.
\nonumber
\end{eqnarray}
Here we have introduced the polarization operator $\hat{\bf P}({\bf r},t)$
of the ultracold ensemble
\begin{equation}
\label{P}
\hat{\bf P}({\bf r},t)
=
{\bf d}
\hat\psi_1^\dagger({\bf r},t)
\hat\psi_2({\bf r},t)
+
H.c.
=
\hat{\bf P}^+({\bf r},t)
+
\hat{\bf P}^-({\bf r},t)
\;,
\end{equation}
and the operator of the microscopic displacement field is given by
\begin{equation}
\label{dis-mic}
\hat {\bf D}_{mic}({\bf r},t)
=
\sum_{{\bf k}\lambda}
i
\sqrt{\frac{2\pi\hbar\omega_k}{V}}
{\bf e}_\lambda
\hat c_{{\bf k}\lambda}
\exp
\left(
      i {\bf k} {\bf r}
\right)
+ H.c.
\end{equation}

We assume that all contact interactions can be neglected.
In order to find the conditions under which this approximation is valid,
it is necessary to estimate the ratio $U_d/U_g$, where $U_d$ and $U_g$
are the mean energies of the dipole-dipole interaction and the ground-state
collisions, respectively. Treating the ground-state collisions in terms
of s-wave scattering and restricting to dipole optical transitions
we get the following inequality\cite{WAL97,KBA99}
\begin{equation}
\label{estimate}
\frac{U_d}{U_g} \gg 37.5 s
\;,
\end{equation}
where $s$ is the saturation parameter of the atomic transition.
Thus, the ground-state collisions are negligible if the saturation parameter
$s$ is of the order of $0.01$ or higher. This shows, that we don't need
very high values of the parameter $s$ in order to neglect contact
interactions.

\section{Heisenberg equations of motion for the atomic and
	 photonic operators}
~~~

Making use of the Hamiltonian (\ref{ham-second-quant}) and the commutation
relations (\ref{komm-equal}) we obtain the following Heisenberg equations
of motion for the atomic and photonic operators:
\begin{eqnarray}
\label{heis}
i \hbar \frac{\partial\hat{\psi}_1({\bf r},t)}{\partial t}
=
&-&
\frac{\hbar^2 \nabla^2}{2m}
\hat{\psi}_1({\bf r},t)
-
{\bf d} {\bf E}_{in}^-({\bf r},t)
\hat{\psi}_2({\bf r},t)
\nonumber\\
&-&
\hat{\psi}_2({\bf r},t)
{\bf d} {\bf E}_{in}^+({\bf r},t)
\nonumber\\
&-&
\hbar \sum_{{\bf k}\lambda} g_{{\bf k}\lambda}^*
\hat{c}^{\dagger}_{{\bf k}\lambda}(t)
\exp
\left(
     - i {\bf k} {\bf r}
\right)
\hat{\psi}_2({\bf r},t)
\nonumber\\
&-&
\hbar \hat{\psi}_2({\bf r},t)
\sum_{{\bf k}\lambda}
g_{{\bf k}\lambda}
\exp
\left(
     i {\bf k} {\bf r}
\right)
\hat{c}_{{\bf k}\lambda}(t)
\;,
\label{heis-a}
\\
i \hbar
\frac{\partial\hat{\psi}_2({\bf r},t)}{\partial t}
=
&-&
\frac{\hbar^2 \nabla^2}{2m}
\hat{\psi}_2({\bf r},t)
+
\hbar \omega_a
\hat{\psi}_2({\bf r},t)
\nonumber\\
&-&
{\bf d} {\bf E}_{in}^-({\bf r},t)
\hat{\psi}_1({\bf r},t)
-
\hat{\psi}_1({\bf r},t)
{\bf d} {\bf E}_{in}^+({\bf r},t)
\nonumber\\
&-&
\hbar
\sum_{{\bf k}\lambda}
g_{{\bf k}\lambda}^*
\hat c^{\dagger}_{{\bf k}\lambda}(t)
\exp
\left(
     - i {\bf k} {\bf r}
\right)
\hat{\psi}_1({\bf r},t)
\nonumber\\
&-&
\hbar \hat{\psi}_1({\bf r},t)
\sum_{{\bf k}\lambda}
g_{{\bf k}\lambda}
\exp
\left(
     i {\bf k} {\bf r}
\right)
\hat c_{{\bf k}\lambda}(t)
\;,
\label{heis-b}
\\
i \hbar \frac{\partial \hat{c}_{{\bf k}\lambda}(t)}{\partial t}
=&&
\hbar
\omega_k \hat{c}_{{\bf k}\lambda}(t) - \hbar g_{{\bf k}\lambda}^*
\int d{\bf r} \exp\left( - i {\bf k} {\bf r} \right)
\nonumber\\
&\times&
\left[
\hat{\psi}_2^\dagger({\bf r},t) \hat{\psi}_1({\bf r},t) +
\hat{\psi}_1^\dagger({\bf r},t) \hat{\psi}_2({\bf r},t)
\right]
\;,
\label{heis-c}
\\
g_{{\bf k}\lambda}
&=&
i \sqrt{\frac{2\pi\omega_k}{\hbar V}} {\bf d}
{\bf e}_\lambda
\;,
\nonumber
\end{eqnarray}
where ${\bf E}_{in}^\pm$ are the positive and negative frequency parts of the
incident classical electric field. The operator products in
Eqs.(\ref{heis-a}),(\ref{heis-b}),(\ref{heis-c}) are taken in normally
ordered form.

The formal solution of (\ref{heis-c}) for the photon operators is
\begin{eqnarray}
\label{solution-photon}
\hat{c}_{{\bf k}\lambda}(t)
&=&
\hat c_{{\bf k}\lambda}(0)
\exp
\left(
     - i \omega_k t
\right)
     \nonumber\\
     &&+ i g_{{\bf k}\lambda}^*
     \int_0^t
   dt'
     \int d{\bf r}'
     \exp
     \left[
	  i \omega_k (t'-t) - i {\bf k} {\bf r}'
     \right]
     \nonumber\\
     &\times&
     \left[
	  \hat \psi_2^\dagger({\bf r}',t')
	  \hat{\psi}_1({\bf r}',t')
	  +
	  \hat{\psi}_1^\dagger({\bf r}',t')
	  \hat{\psi}_2({\bf r}',t')
     \right]
\;,
\end{eqnarray}
where the first term $\hat{c}_{{\bf k}\lambda}(0)$ refers to the free-space
photon field and the second one goes back to the interaction with the atoms.

To study the back reaction of the photons on matter we insert
(\ref{solution-photon}) in (\ref{heis-a}) and (\ref{heis-b}).
In the rotating wave approximation we obtain the dynamical equations
for the operators of the two matter fields
\begin{eqnarray}
\label{matter}
i \hbar \frac{\partial\hat{\psi}_1({\bf r},t)}{\partial t}
&=&
-
\frac{\hbar^2 \nabla^2}{2m} \hat{\psi}_1({\bf r},t)
-
{\bf d} \hat{{\bf E}}_{loc}^-({\bf r},t)
\hat{\psi}_2({\bf r},t)
\;,
\label{matter-a}
\\
i \hbar \frac{\partial\hat{\psi}_2({\bf r},t)}{\partial t}
&=&
-
\frac{\hbar^2 \nabla^2}{2m} \hat{\psi}_2({\bf r},t)
+
\hbar
\left(
     \omega_a + \delta - i \gamma/2
\right)
\hat\psi_2({\bf r},t)
\nonumber\\
&&-
\hat\psi_1({\bf r},t)
{\bf d} \hat{\bf E}_{loc}^+({\bf r},t)
\;,
\label{matter-b}
\end{eqnarray}
where $\delta$ and $\gamma$ are the Lamb shift and the spontaneous
emission rate of a single atom in free space, respectively,
and we have introduced the operator of the local electric field
\begin{eqnarray}
\label{def-e-local}
\hat{{\bf E}}_{loc}^+({\bf r},t)
=
&&
{\bf E}_{in}^+({\bf r},t)
\nonumber\\
&&+
i
\sum_{{\bf k}\lambda}
\sqrt{\frac{2\pi\hbar\omega_k}{V}}
{\bf e}_\lambda
\hat{c}_{{\bf k}\lambda}(0)
\exp
\left(
     i {\bf k} {\bf r} - i \omega_k t
\right)
\nonumber\\
&&
+
\int
d{\bf r}'
\nabla \times \nabla \times
\frac { \hat{\bf P}^+ \left( {\bf r}',t-R/c \right)}{R}
\;,
\end{eqnarray}
and $\nabla\times$ refers to the point ${\bf r}$.
Note that in eq.(\ref{def-e-local}) a small volume around the
observation point ${\bf r}$ is excluded from the integration.

Eq. (\ref{def-e-local}) shows that $\hat{{\bf E}}_{loc}^\pm({\bf r},t)$ 
is the superposition of the incident field ${\bf E}_{in}^\pm({\bf r},t)$, 
vacuum fluctuations of the photon field, and the electric field
radiated by all other atoms.
As one should expect it is this local field, which
drives the inner atomic transition in (\ref{matter-a}),
(\ref{matter-b}).

Equations (\ref{matter-a}), (\ref{matter-b}) can be regarded as
an atom-optical
analogue of the optical Bloch equations\cite{ALL78,BOW93}.
They describe the dynamical evolution of the second quantized matter
in the field of electromagnetic radiation.

\section{Lorentz-Lorenz relation and the system of
Maxwell-Bloch equations in atom optics}
~~~

\subsection{Local-field correction}
~~~

Treating the quantum gas as a medium, it is important to establish a
connection to the observables of the quantum electrodynamics in media. 
Having an application to an ultracold gas in mind, we have to relate 
the local field $\hat{{\bf E}}_{loc}({\bf r},t)$ to the {\it macroscopic
(or mean) field} $\hat{{\bf E}}_{mac}({\bf r},t)$ which is obtained by 
averaging in space over a region which contains a great number of atoms. 
For a very low density ultracold gas the region on which the
macroscopic field is averaged may contain actually fewer than one atom,
yet the averaging technique is still valid due to atomic
delocalization.

We generalize the idea of Lorentz as it is described
in\cite{BOW93,BOR68,JAC75eng} and assume the local field to be a sum of
several contributions:
\begin{equation}
\label{L-method}
\hat {\bf E}_{loc}({\bf r},t) =
\hat {\bf E}_{near}({\bf r},t) +
\hat {\bf E}_{mac}({\bf r},t) -
\hat {\bf E}_P({\bf r},t)
\;.
\end{equation}
Let us consider a macroscopically small, but microscopically large volume
${\cal V}$ around ${\bf r}$. $\hat {\bf E}_{near}$ is a contribution from
the dipoles within ${\cal V}$. $\hat {\bf E}_{mac}$ is the macroscopic
field obtained from the field of all dipoles in the medium.
Because the dipoles in ${\cal V}$ are already taken into account in
$\hat {\bf E}_{near}$, we have to subtract from $\hat {\bf E}_{mac}$
the contribution $\hat {\bf E}_P$ of these dipoles as it is obtained in
the averaged continuum approximation. $\hat {\bf E}_P$ is therefore
a macroscopic quantity. It can be related to the polarization,
which is an averaged quantity too, according to
\begin{equation}
\label{EP}
\hat {\bf E}_P({\bf r},t) = - \frac{4\pi}{3} \hat {\bf P}({\bf r},t)
\;.
\end{equation}
For a novel derivation see\cite{BOW93}. In the following we restrict
to the purely classical local-field correction which corresponds to
a vanishing $\hat {\bf E}_{near}$ \cite{BOW93,BOR68,JAC75eng}.
Thus, we see that, because of the influence of the near dipole-dipole
interactions, the local field is obtained from the macroscopic field
in adding the {\it local-field correction} \cite{BOW93,BOR68,JAC75eng}.
\begin{equation}
\label{local-corr}
\hat{{\bf E}}_{loc}^\pm({\bf r},t) =
\hat{{\bf E}}_{mac}^\pm({\bf r},t) +
\frac{4\pi}{3}\hat{{\bf P}}^\pm({\bf r},t)
\;.
\end{equation}
This equation is also often called the {\it Lorentz-Lorenz relation}. The
propagation of the operator of the macroscopic field is thereby given
by Maxwell wave equation for a charge-free and current-free
polarization medium.

\subsection{Nonlinear matter equation}
~~~

In the next step we decouple the dynamical equations for the two matter field
operators. We substitute (\ref{local-corr}) in (\ref{matter-a}) and
(\ref{matter-b}) and pass to the reference frame rotating with the frequency
$\omega_L$ of the incident field which is assumed to be monochromatic
\begin{eqnarray}
\label{transform-rot}
\hat{{\bf E}}_{mac}^+({\bf r},t)
&=&
\hat{\bf \cal E}_{mac}^+({\bf r})
\exp
\left(
    - i \omega_L t
\right)
\;,
\nonumber\\
\hat{\psi}_2({\bf r},t)
&=&
\hat{\phi}_2({\bf r},t)
\exp
\left(
     - i \omega_L t
\right)
\;,
\end{eqnarray}
to obtain
\begin{eqnarray}
\label{nonlinear}
i \hbar
\frac{\partial \hat \psi_1}{\partial t}
&=&
-
\frac{\hbar^2 \nabla^2}{2m}
\hat \psi_1
-
\frac{\hbar}{2}
\hat{\Omega}^-({\bf r})
\hat \phi_2
-
\frac{4\pi}{3}
d^2
\hat \phi_2^\dagger
\hat \psi_1 \hat \phi_2
\;,
\label{nonlinear-a}
\\
i \hbar
\frac{\partial\hat{\phi}_2}{\partial t}
&=&
-
\frac{\hbar^2 \nabla^2}{2m}
\hat{\phi}_2
-
\frac{\hbar}{2}
\hat{\psi}_1
\hat{\Omega}^+({\bf r})
-
\frac{4\pi}{3} d^2
\hat{\psi}_1
\hat{\psi}_1^\dagger
\hat{\phi}_2
\nonumber\\
&&
-
\hbar
\left( \Delta + i \gamma/2 \right) \hat{\phi}_2
\;,
\label{nonlinear-b}
\end{eqnarray}
with the detuning $\Delta=\omega_L-\omega_a-\delta$. The position dependent
Rabi frequency
$\hat{\Omega}^\pm({\bf r})=2{\bf d}\hat{\bf \cal E}_{mac}^\pm({\bf r})/\hbar$
is related to the macroscopic electric field.

We assume sufficiently large detuning such that the contribution of the
spontaneous emission is small. We may therefore apply
{\it the adiabatic approximation} to (\ref{nonlinear-b}),
which gives
\begin{equation}
\label{adiabatic-sol}
\hat{\phi}_2({\bf r},t) =
-
\frac
{\hat{\Omega}^+({\bf r}) \hat{\psi}_1({\bf r},t)}
{ 2\left[ \hat{\Delta}_l({\bf r},t) + i \gamma/2 \right] }
\;,
\end{equation}
and eliminate the field $\hat{\psi}_2$ from our scheme. The position
dependent {\it local detuning} introduced in
\begin{equation}
\label{def-local-det}
\hat{\Delta}_l({\bf r},t) =
\Delta +
\frac{4\pi}{3\hbar} d^2
\hat{\psi}_1^\dagger({\bf r},t)
\hat{\psi}_1({\bf r},t)
\end{equation}
is an operator depending on the density operator of the ground state. We
interpret this as a shift of the internal energy levels which may increase
or decrease with increasing density depending on the sign of the detuning
$\Delta$. Analogous frequency shifts can be found in nonlinear optics
(see, for instance, \cite{BOW93,CB96} and references therein).

Because we are mainly interested in atom optical problems and want
to study the coherent evolution of the center-of-mass motion of the gas,
we shall neglect spontaneous emission. This is valid for situations
where the absolute value of the local detuning is much bigger than
the spontaneous emission rate $\left|\Delta_l\right| \gg \gamma$.
In order to do this we drop in the following the vacuum fluctuations
and the spontaneous emission rate $\gamma$ from our equations.

Then substituting (\ref{def-local-det}) in
(\ref{nonlinear-a}) we obtain as the intended result the nonlinear dynamical
equation for the matter field operator $\hat{\psi}_1({\bf r},t)$
\begin{equation}
\label{nonlinear-equation}
i\hbar
\frac{\partial \hat{\psi}_1({\bf r},t)}{\partial t}
=
\left\{
     -\frac
     {\hbar^2\nabla^2}{2m} +\frac{\hbar}{4} \frac{\Delta}
     {\hat{\Delta}_l^2({\bf r},t)}
     \left|
	 \hat{\Omega}^+({\bf r})
     \right|^2
\right\}
\hat{\psi}_1({\bf r},t)
\;.
\end{equation}
It couples to the macroscopic electric field via the Rabi frequency
$\hat{\Omega}^\pm({\bf r})$.

On the assumptions stated in the Section I, equation
(\ref{nonlinear-equation})
is general. No approximation apart from the adiabatic approximation
and the rotating wave approximation has been made.
The dynamical dipole-dipole interaction mediated through
the exchange of photons is completely contained.
The nonlinearity goes back to the density
dependent local detuning $\hat{\Delta}_l({\bf r},t)$ of
eq.~(\ref{def-local-det}). For increasing density and positive detuning
$\Delta$ the local detuning grows and correspondingly the nonlinear term
in (\ref{nonlinear-equation}) representing the coupling to the macroscopic
electric field becomes smaller. On the other hand, for negative detuning
the absolute value of the local detuning decreases with the increase
of the density and the nonlinearity becomes greater.

\subsection{Microscopic and macroscopic Maxwell equations}
~~~
In the multipolar formulation of QED the operator of the microscopic
displacement field is given by eq.(\ref{dis-mic}) and
the operator of the magnetic field has the form
\begin{eqnarray}
\label{b-mic}
\hat{\bf B}_{mic} ({\bf r},t)
=
&&
i
\sum_{{\bf k}\lambda}
\sqrt{\frac{2\pi\hbar\omega_k}{V}}
\frac{\omega_k}{c}
\left(
     {\bf k} \times {\bf e}_\lambda
\right)
   \hat c_{{\bf k}\lambda}(t)
   \exp
   \left(
       i {\bf k} {\bf r}
   \right)
\nonumber\\
&&
   +
H.c.
\end{eqnarray}

Calculating time and spatial derivatives of the operators (\ref{dis-mic}),
(\ref{b-mic}) we get microscopic Maxwell equations
\begin{eqnarray}
\label{maxwell-mic}
\nabla \times \hat{\bf E}_{mic}
&=&
- \frac{1}{c} \frac{\partial}{\partial t} \hat{\bf B}_{mic}
,
\quad
div \hat{\bf B}_{mic} = 0
,
\label{maxwell-mic-a}
\nonumber\\
\nabla \times \hat{\bf H}_{mic}
&=&
\frac{1}{c} \frac{\partial}{\partial t} \hat{\bf D}_{mic}
,
\quad
div \hat{\bf D}_{mic} = 0
,
\label{maxwell-mic-b}
\end{eqnarray}
where
$\hat{\bf D}_{mic} = \hat{\bf E}_{mic} + 4\pi\hat{\bf P}$
and $\hat{\bf B}_{mic}=\hat{\bf H}_{mic}$.
The same equations were obtained by G.Juze\-liunas~\cite{Juz97} in
a polariton model of a dielectric medium interacting with light.

In order to get the macroscopic Maxwell equations we have to average
eqs.(\ref{maxwell-mic}) over the microscopically large, but
macroscopically small volume of space. However, in the case of
ultracold atomic gases due to the atomic delocalization any point
of space contains contributions from all the atoms. In this sense
the microscopic field is already averaged and we may simply replace
the microscopic fields by their macroscopic analogues.

Assuming that the spatial variations of the atomic density are not rapid
and taking into account the adiabatic solution (\ref{adiabatic-sol}),
one can rewrite the system of macroscopic Maxwell equations in the form
of the wave equation
\begin{equation}
\label{maxwell-media}
\nabla^2 \hat{\bf \cal E}_{mac}^\pm  +
k_L^2 \hat{n}^2 \hat{\bf \cal E}_{mac}^\pm = 0
\;,
\end{equation}
where the refractive index is given by the Clausius-Mossotti relation
\begin{equation}
\label{clausius}
\hat{n}^2
=
\frac
{
  1 +
  \frac{8\pi}{3}
  \alpha \hat{\psi}_1^\dagger \hat{\psi}_1
}
{1 - \frac{4\pi}{3} \alpha \hat{\psi}_1^\dagger \hat{\psi}_1}
\;.
\end{equation}

Eqs.(\ref{nonlinear-equation}),(\ref{maxwell-media}) and
(\ref{clausius}) constitute the general system of Maxwell-Bloch
equations, which can be employed for the solution of different
problems in nonlinear atom optics.
In our paper\cite{KBA99} we have shown that
the equations which are usually used for the description of an ultracold
ensemble put in an external laser field can be derived from
(\ref{nonlinear-equation}),(\ref{maxwell-media}) and
(\ref{clausius}) assuming that the density of atoms is
a small parameter.

\subsection{Equal-time commutation relations for the operators
of the electromagnetic field}
~~~

The operators of the local electric and magnetic field strength have
the following form:
\begin{eqnarray}
\label{EB-loc}
\hat{\bf E}_{loc} ({\bf r},t)
&=&
i
\sum_{{\bf k}\lambda}
\sqrt{\frac{2\pi\hbar\omega_k}{V}}{\bf e}_\lambda
   \hat c_{{\bf k}\lambda}(t)
   \exp
   \left(
       i {\bf k} {\bf r}
   \right)
   +
H.c.
\nonumber\\
&&
- \frac{8\pi}{3} \hat {\bf P} ({\bf r},t)
\label{EB-loc-a}
\\
\hat{\bf B}_{loc} ({\bf r},t)
&=&
i
\sum_{{\bf k}\lambda}
\sqrt{\frac{2\pi\hbar\omega_k}{V}}
\frac{\omega_k}{c}
\left(
     {\bf k} \times {\bf e}_\lambda
\right)
   \hat c_{{\bf k}\lambda}(t)
   \exp
   \left(
       i {\bf k} {\bf r}
   \right)
\nonumber\\
&&
   +
H.c.
\label{EB-loc-b}
\end{eqnarray}

Taking into account the commutation relations for the photons and
matter-field operators
\begin{eqnarray}
\label{com-m-ph}
\left[
     \hat c_{{\bf k}\lambda}(t), \hat c_{{\bf k}\lambda}^\dagger(t)
\right]
&=&
1
,
\nonumber\\
\left[
     \hat c_{{\bf k}\lambda}(t), \hat \psi_{1,2}(t)
\right]
&=&
\left[
     \hat c_{{\bf k}\lambda}(t), \hat \psi_{1,2}^\dagger(t)
\right]
=
0
,
\end{eqnarray}
we obtain the following equal-time commutation relations for the operators
of the local electromagnetic field
\begin{eqnarray}
\label{com-loc}
&&
\left[
    \hat{\bf E}_{loc}^m({\bf r}_1,t) , \hat{\bf E}_{loc}^n({\bf r}_2,t)
\right]
=
\left(
\frac{8\pi}{3}
\right)^2
d^m d^n
\delta({\bf r}_1-{\bf r}_2)
\label{com-loc-a}
\nonumber\\
&&
\times
\left[
    \hat\psi_1^\dagger({\bf r}_1,t) \hat\psi_1({\bf r}_2,t)
    -
    \hat\psi_1^\dagger({\bf r}_2,t) \hat\psi_1({\bf r}_1,t)
\right.
\nonumber\\
&&
\hspace{1cm}
\left.
    +
    \hat\psi_2^\dagger({\bf r}_1,t) \hat\psi_2({\bf r}_2,t)
    -
    \hat\psi_2^\dagger({\bf r}_2,t) \hat\psi_2({\bf r}_1,t)
\right]
\;,
\\
&&
\left[
    \hat{\bf E}_{loc}^m({\bf r}_1,t) , \hat{\bf B}_{loc}^n({\bf r}_2,t)
\right]
=
- i 4\pi \hbar c
\sum_l
\varepsilon_{mnl}
\frac{\partial}{\partial R_l} \delta({\bf R})
.
\label{com-loc-b}
\end{eqnarray}
In eq.(\ref{com-loc-a}) the $\delta$-function vanishes,
if ${\bf r}_1\not={\bf r}_2$. If ${\bf r}_1 = {\bf r}_2$, the term in
square brackets in the r.h.s. of eq.(\ref{com-loc-a}) vanishes.
Therefore, the commutator (\ref{com-loc-a}) vanishes for any
${\bf r}_1$ and ${\bf r}_2$, and we have the canonical commutation 
relations for the operators of the local electromagnetic field.

Starting from the Lorentz-Lorenz relation (\ref{local-corr}) and 
making use of the equal-time commutation relations
(\ref{com-loc-a}), (\ref{com-loc-b})
we get the canonical commutation relations for the operators of the 
macroscopic electromagnetic field:
\begin{eqnarray}
\label{com-mac}
&&
\left[
    \hat{\bf E}_{mac}^m({\bf r}_1,t) , \hat{\bf E}_{mac}^n({\bf r}_2,t)
\right]
=
0
\;,
\label{com-mac-a}
\\
&&
\left[
    \hat{\bf E}_{mac}^m({\bf r}_1,t) , \hat{\bf B}_{mac}^n({\bf r}_2,t)
\right]
=
\nonumber\\
&&
\hspace{2cm}
- i 4\pi \hbar c
\sum_l 
\varepsilon_{mnl}
\frac{\partial}{\partial R_l} \delta({\bf R})
\;.
\label{com-mac-b}
\end{eqnarray}

The commutation relations (\ref{com-loc-a})-(\ref{com-mac-b}) show that
our theory is consistent with the general principles of the canonical
quantization of the electromagnetic field in a medium.

\section{Conclusion}
~~~

We have treated the interaction of dense Bose and Fermi gases
with light in the microscopic approach.
Our results
are not limited to low densities and include the dipole-dipole interaction
consistently.
The present analysis is concentrated on the common features
of the interaction of Bose  and Fermi gases with the radiation field.
The differences appear when we consider correlation functions which
reflect Bose or Fermi statistics. The effects caused by the statistics
where considered in \cite{MOR95,RJ,RJcm}, where it was shown that
the refractive index contains additional terms. However, these statistical
effects are not very important because the leading term is governed
by the Clausius-Mossotti formula.

\section*{Acknowledgments}
~~~

This work has been supported by the Deutsche For\-schungsgemeinschaft
and the Optikzentrum Konstanz. One of us (K.V.K.) would like to
thank the members of the AG Audretsch at the University of Kon\-stanz
for many interesting discussions and kind hospitality. Valuable
discussions with G.Hegerfeldt,
D.-G.Welsch, S.Scheel, G.Juzeliunas, J.Ruostekoski, and the group
of Prof.W.Schleich at Ulm university are greatfully
acknowledged. Special thanks to C.M.Bowden for his interest to our work.



\begin{thebibliography}{99}

\bibitem{MOR95}
Morice, O., Castin, Y., and Dalibard, J.,
1995, {\it Phys.Rev.A}, {\bf 51}, 3896.

\bibitem{RJ}
Ruostekoski, J., and Javanainen, J.,
1997, {\it Phys.Rev.A}, {\bf 55}, 513.

\bibitem{RJcm}
Ruostekoski, J., and Javanainen, J.,
1999, {\it cond-mat/9902172}.

\bibitem{LEN93}
Lenz, G., Meystre, P., and Wright, E.,
1993, {\it Phys.Rev.Lett.}, {\bf 71}, 3271.

\bibitem{ZHA94a}
Zhang, W., and Walls, D.,
1994, {\it Phys.Rev.A}, {\bf 49}, 3799.

\bibitem{LEN94}
Lenz, G., Meystre, P., and Wright, E.,
1994, {\it Phys.Rev.A}, {\bf 50}, 1681.

\bibitem{CAS95}
Castin, Y., and M{\o}lmer, K.,
1995, {\it Phys.Rev.A}, {\bf 51},  R3426.

\bibitem{WAL97}
Wallis, H.,
1997, {\it Phys.Rev.A}, {\bf 56},  2060.

\bibitem{KBA99}
Krutitsky, K.V., Burgbacher, F., and Audretsch, J.,
1999, {\it Phys.Rev.A}, {\bf 59}, 1517.

\bibitem{COH89}
Cohen-Tannoudji, C., Dupont-Roc, J., and Grynberg, G.,
1989, {\it Photons and Atoms: Introduction to Quantum Electrodynamics}
(New York: Wiley \& Sons).

\bibitem{BOW93}
Bowden, C.M., and Dowling, J.P.,
1993, {\it Phys.Rev.A}, {\bf 47}, 1247.

\bibitem{BOR68}
Born, M., and Wolf, E.,
1970, {\it Principles of Optics}
(New York: Pergamon).

\bibitem{JAC75eng}
Jackson, J.,
1975, {\it Classical Electrodynamics}
(New York: Wiley).

\bibitem{ALL78}
Allen, L., and Eberly, J.,
1978, {\it Optical Resonance and Two-Level Atoms}
(New York: Wiley).

\bibitem{CB96}
Crenshaw, M.E., and Bowden, C.M.,
1996, {\it Phys.Rev.A}, {\bf 53}, 1139.

\bibitem{Juz97}
Juzeliunas, G.,
1997, {\it Phys.Rev.A}, {\bf 55}, 929.

\end{thebibliography}
\end{document}